\documentclass[12pt]{article}

\input epsf
\usepackage{amssymb}
\usepackage[dvips]{graphicx}
\usepackage{amsmath,amscd}

\def\dj{\hbox{d\kern-0.347em \vrule width 0.3em height 1.252ex depth
-1.21ex \kern 0.051em}}

\numberwithin{equation}{section}

\begin{document}

\setlength{\oddsidemargin}{0cm}
\setlength{\baselineskip}{7mm}


\thispagestyle{empty}
\setcounter{page}{0}

\begin{flushright}

\end{flushright}

\vspace*{1cm}

\begin{center}
{\bf \Large Twisted invariances of noncommutative gauge
theories}

\vspace*{1cm}

\'Alvaro Due\~nas-Vidal\footnote{\tt 
adv@usal.es}
and Miguel A. V\'azquez-Mozo\footnote{\tt 
Miguel.Vazquez-Mozo@cern.ch}

\end{center}

\vspace*{0.0cm}

\begin{center}
  
 {\sl Departamento de F\'{\i}sica Fundamental,
 Universidad de Salamanca \\ 
 Plaza de la Merced s/n,
 E-37008 Salamanca, Spain
  }

\end{center}

\vspace*{3cm}

\centerline{\bf \large Abstract}

We study noncommutative deformations of Yang-Mills theories and show that these theories 
admit a infinite, continuous family of 
twisted star-gauge invariances. This family interpolates continuously between 
star-gauge and twisted gauge transformations. The possible physical r\^ole of these start-twisted 
invariances is discussed. 

\noindent

\newpage

\section{Introduction}

Noncommutative deformations of field theories have provided interesting workbenches where some
properties of Quantum Field Theory  can be probed (for reviews see \cite{reviews}). Of particular interest in 
this field has been the study of gauge theories on noncommutative spaces. Apart from many other interesting features,
these theories describe certain low-energy limit of string theory in the presence of a constant $B$-field background \cite{seiberg_witten}.

Although very interesting from a mathematical physics point of view, noncommutative deformation of Yang-Mills theories 
seem to have a limited phenomenological interest. In general, the deformation of the gauge transformations
force the gauge fields to take values in the universal enveloping algebra of the gauge group \cite{uea}. 
In particular, it can be seen that the only gauge group for which the gauge transformations close is U($N$) \cite{gauge-groups,no-go}, 
thus excluding the phenomenologically more interesting special unitary groups (see however \cite{other_groups} for some proposals
to realize nonunitary groups in noncommutative geometry). In addition to this, 
the theory suffers from instabilities at the quantum level. A one loop calculation of the dispersion
relation for the noncommutative photon shows that it diverges at low momentum \cite{tachyons}. 
This instability can be removed by embedding the gauge theory into noncommutative $\mathcal{N}=4$ super-Yang-Mills theory 
at high energy, but at the price of introducing a severe fine tuning of the supersymmetry breaking scale \cite{LAG-VM}.
Alternatively, lattice studies of noncommutative U(1) gauge theories have shown \cite{lattice} that the tachyonic instability
can be eliminated in a new nonperturbative phase of the theory characterized by the breaking of translation invariance.

In the construction of Yang-Mills theories on noncommutative spaces presented in \cite{seiberg_witten} star-gauge transformations 
play a central r\^ole as the  true gauge symmetry of the deformed theory \cite{star-gauge}. This deformed 
symmetry acts in the standard way by (nonlocal) transformation of the fields. On the other hand, in 
\cite{vassilevich,twisted_gauge} it was pointed out that the action of noncommutative Yang-Mills is also invariant under standard, i.e., commutative,
gauge transformations provided the Leibniz rule is twisted accordingly. Although in this case the  
transformations are consistent for any gauge group, the equations of motion of the
theory force now the gauge fields to take values on the universal enveloping algebra of the gauge group \cite{twisted_gauge}. 
The study of these type of theories has attracted considerable attention \cite{twisted}.

The extra terms appearing in the twisted Leibniz rule in this type of theories can be understood 
as due to a transformation of the star-product itself
under gauge transformations \cite{LAG-M-VM}. From this point of view twisted gauge transformations are not standard, {\em bona-fide} 
transformations since they involve not only the transformation of fields but of the product operation as well. This prevents a direct
application of the standard procedures to obtain Noether currents and/or Ward identities associated with these symmetries.  It is important
to keep in mind that twisted gauge theories are also invariant under the corresponding star-gauge transformation, which is a standard
symmetry of the theory acting only on fields. In \cite{LAG-M-VM} it
was argued that star-gauge transformations play a custodial r\^ole in guaranteeing the existence of conserved current and Ward identities.
This point of view was further supported in \cite{giller_et_al}, where it was argued that the consistency of the twisted gauge
theory requires the presence of the custodian star-gauge symmetry. 

The crucial point in the construction of noncommutative twisted gauge theories is the realization that the product used in writing the action
and the product involved in the gauge transformations does not have to be the same if other conditions like the Leibniz rule
are relaxed. In the case of Refs. \cite{vassilevich,twisted_gauge} gauge transformations act through the 
ordinary, commutative, product, whereas the action
is constructed in terms of the star-product. 
In this note we show that this construction  can be generalized in such a way that noncommutative gauge theories can be shown to be
invariant under star-gauge transformations defined with {\sl any} noncommutative parameter, with the appropriate twist of the Leibniz rule. 
This family of invariances continuously interpolate between star-gauge symmetry and twisted gauge transformations defined
in terms of the standard commutative product.

\section{Star-twisted gauge transformations: Heuristic derivation}

\paragraph{Pure noncommutative Yang-Mills.}
Let us consider the algebra $\mathcal{A}$ of functions on $\mathbb{R}^{d}$ and the Groenewold-Moyal 
star-product between elements of this algebra defined as
\begin{eqnarray}
f(x)\star_{\theta} g(x) \equiv f(x) \exp\left[{{i\over 2}\theta^{\mu\nu}\overleftarrow{\partial}_{\mu}\overrightarrow{\partial}_{\nu}}
\right]g(x).
\label{moyal}
\end{eqnarray}
For the sake of clarity, here and in the following we always denote explicitly the noncommutativity parameter used in the
definition of the star-product.
Pure gauge theories on noncommutative spaces can be constructed in terms of the previous noncommutative product by
\cite{reviews}
\begin{eqnarray}
S=-{1\over 2g^{2}}\int d^{d}x\,{\rm tr\,}\Big(F_{\mu\nu}\star_{\theta}F^{\mu\nu}\Big), \hspace*{1cm} 
F_{\mu\nu}=\partial_{\mu}A_{\nu}-\partial_{\nu}A_{\mu}-i[A_{\mu},A_{\nu}]_{\theta},
\label{action}
\end{eqnarray} 
where we have used the obvious notation $[A,B]_{\theta}\equiv A\star_{\theta}B-B\star_{\theta} A$. 
This action is invariant under
the star-gauge symmetry
\begin{eqnarray}
\delta_{\varepsilon}^{\theta}A_{\mu}=\partial_{\mu}\varepsilon+i[\varepsilon,A_{\mu}]_{\theta}.
\label{stargauge1}
\end{eqnarray}
Because of this deformed gauge transformation the gauge group has to be restricted to U($N$). For any other gauge group $G$, 
Eq. (\ref{stargauge1}) forces the gauge 
field $A_{\mu}$ has to take values on the universal enveloping algebra of the Lie algebra of $G$. Here we 
confine our analysis to theories with gauge group U($N$).

The idea of Ref. \cite{vassilevich,twisted_gauge} is that the action (\ref{action}) can also be made invariant under standard 
(undeformed) gauge 
transformations
\begin{eqnarray}
\delta_{\varepsilon}^{0}A_{\mu}=\partial_{\mu}\varepsilon+i[\varepsilon,A_{\mu}]_{0}\equiv \partial_{\mu}
\varepsilon
+i\Big(\varepsilon\cdot A_{\mu}-A_{\mu}\cdot \varepsilon\Big),
\end{eqnarray}
provided the action of the transformations on the products of fields is changed appropriately
\begin{eqnarray}
\delta_{\varepsilon}^{0}\Big(A_{\mu}\star_{\theta} A_{\nu}\Big) \hspace*{13cm}  \\
\hspace*{1cm}=\sum_{n=0}^{\infty}{(-i/2)^{n}\over n!}\theta^{\alpha_{1}\beta_{1}}
\theta^{\alpha_{2}\beta_{2}}\ldots\theta^{\alpha_{n}\beta_{n}}
\Big\{\Big([\partial_{\alpha_{1}},[\partial_{\alpha_{2}},\ldots[\partial_{\alpha_{n}},\delta_{\varepsilon}^{0}]\ldots]]A_{\mu}\Big)\star_{\theta}
\Big(
\partial_{\beta_{1}}\partial_{\beta_{2}}\ldots
\partial_{\beta_{n}}A_{\nu}\Big) \nonumber \\
+\Big(\partial_{\alpha_{1}}\partial_{\alpha_{2}}\ldots
\partial_{\alpha_{n}}A_{\mu}\Big)\star_{\theta}
\Big([\partial_{\beta_{1}},[\partial_{\beta_{2}},\ldots[\partial_{\beta_{n}},
\delta_{\varepsilon}^{0}]\ldots]]A_{\nu}\Big)\Big\}.
\nonumber
\end{eqnarray}
In this series the term $n=0$ gives the standard Leibniz rule which is corrected by an infinite number of terms with arbitrary number of
derivatives. This transformation of the product of two gauge fields implies that the field strength transforms as
\begin{eqnarray}
\delta_{\varepsilon}^{0}F_{\mu\nu}=[i\varepsilon,F_{\mu\nu}]_{0}\equiv i\Big(\varepsilon\cdot F_{\mu\nu}-F_{\mu\nu}\cdot \varepsilon\Big).
\end{eqnarray}
This transformation, together with the twisted Leibniz rule, 
guarantees the invariance of the action under twisted gauge transformations.

Let us now go back to the action (\ref{action}) but consider a star-gauge transformation with parameter $\theta'{}^{\mu\nu}\neq
\theta^{\mu\nu}$
\begin{eqnarray}
\delta_{\varepsilon}^{\theta'}A_{\mu}=\partial_{\mu}\varepsilon+i[\varepsilon,A_{\mu}]_{\theta'}.
\label{star-twisted}
\end{eqnarray}
The variation of the field strength $F_{\mu\nu}$ in Eq. (\ref{action}) under this transformation can be written as
\begin{eqnarray}
\delta_{\varepsilon}^{\theta'}F_{\mu\nu}=\ [i\varepsilon,\partial_{\mu}A_{\nu}-\partial_{\nu}A_{\mu}]_{\theta'}
+i[\partial_{\mu}\varepsilon,A_{\nu}]_
{\theta'}
-i[\partial_{\nu}\varepsilon,A_{\mu}]_{\theta'}-i\delta_{\varepsilon}^{\theta'}[A_{\mu},A_{\nu}]_{\theta}.
\end{eqnarray}
In order to evaluate the last term explicitly we need to compute the action of the $\theta'$-star gauge transformation on the 
$\theta$-star product, $\delta_{\varepsilon}^{\theta'}(A_{\mu}\star_{\theta}A_{\nu})$. For this 
we use the deformed Leibniz rule
\begin{eqnarray}
\delta_{\varepsilon}^{\theta'}\Big(A_{\mu}\star_{\theta} A_{\nu}\Big)
&=&\sum_{n=0}^{\infty}{(-i/2)^{n}\over n!}(\theta^{\alpha_{1}\beta_{1}}-\theta'{}^{\alpha_{1}\beta_{1}})
(\theta^{\alpha_{2}\beta_{2}}-\theta'{}^{\alpha_{2}\beta_{2}})\ldots(\theta^{\alpha_{n}\beta_{n}}-\theta'{}^{\alpha_{n}\beta_{n}})
\nonumber \\
&\times &
\Big\{\Big([\partial_{\alpha_{1}},[\partial_{\alpha_{2}},\ldots[\partial_{\alpha_{n}},\delta_{\varepsilon}^{\theta'}]\ldots]]A_{\mu}\Big)
\star_{\theta}\Big(
\partial_{\beta_{1}}\partial_{\beta_{2}}\ldots
\partial_{\beta_{n}}A_{\nu}\Big)  \label{modAA}\\
& & \,\,\,+\,\,\,\Big(\partial_{\alpha_{1}}\partial_{\alpha_{2}}\ldots
\partial_{\alpha_{n}}A_{\mu}\Big)\star_{\theta}
\Big([\partial_{\beta_{1}},[\partial_{\beta_{2}},\ldots[\partial_{\beta_{n}},\delta_{\varepsilon}^{\theta'}]\ldots]]A_{\nu}\Big)\Big\}.
\nonumber
\end{eqnarray}  
After a tedious but straightforward calculation one arrives at
\begin{eqnarray}
\delta_{\varepsilon}^{\theta'}\Big(A_{\mu}\star_{\theta} A_{\nu}\Big)=(\partial_{\mu}\varepsilon)\star_{\theta'}A_{\nu}+
A_{\mu}\star_{\theta'}(\partial_{\nu}\varepsilon)+[i\varepsilon,A_{\mu}\star_{\theta}A_{\nu}]_{\theta'}.
\end{eqnarray}
In getting this expression we have to use two identities valid for any pair of functions $f(x)$, $g(x)$.
The first equality is
\begin{eqnarray}
{}[f,\partial_{\mu_{1}}\ldots\partial_{\mu_{n}}g]_{\theta'} &=& 
\sum_{k=0}^{n} (-1)^{n-k}
\partial_{(\mu_{1}}\ldots\partial_{\mu_{k}}[\partial_{\mu_{k+1}}\ldots\partial_{\mu_{n})}f,g]_{\theta'},
\label{commutators}
\end{eqnarray}
where the parenthesis indicates the symmetrization
\begin{eqnarray}
\partial_{(\mu_{1}}\ldots\partial_{\mu_{k}}[\partial_{\mu_{k+1}}\ldots\partial_{\mu_{n})}f,g]_{\theta'}
\equiv \sum_{\sigma\in S_{n}}{1\over k!(n-k)!}
\partial_{\sigma(\mu_{1}}\ldots\partial_{\mu_{k}}[\partial_{\mu_{k+1}}\ldots\partial_{\mu_{n})}f,g]_{\theta'}
\end{eqnarray}
with $S_{n}$ the permutation group of $n$ elements. The second crucial identity provides a way to rewrite the
$\theta$-star product in terms of a series of $\theta'$-star products
\begin{eqnarray}
f\star_{\theta} g=\sum_{n=0}^{\infty}{(-i/2)^{n}\over n!}(\theta'{}^{\mu_{1}\nu_{1}}-\theta^{\mu_{1}\nu_{1}})\ldots
(\theta'{}^{\mu_{n}\nu_{n}}-\theta^{\mu_{n}\nu_{n}})(\partial_{\mu_{1}}\ldots\partial_{\mu_{n}}f)\star_{\theta'}
(\partial_{\nu_{1}}\ldots\partial_{\nu_{n}}g).
\label{crucial_identity}
\end{eqnarray}
The proof of this equation is given in the Appendix. Notice that for $\theta'{}^{\mu\nu}=0$ it reduces
to the very definition of the Groenewold-Moyal star product.

Given the deformed Leibniz rule (\ref{modAA}) we find that the transformation of the field strength is given by
\begin{eqnarray}
\delta_{\varepsilon}^{\theta'}F_{\mu\nu}=
\Big[i\varepsilon,\partial_{\mu}A_{\nu}-\partial_{\nu}A_{\mu}-i[A_{\mu},A_{\nu}]_{\theta}\Big]_{\theta'}
=[i\varepsilon,F_{\mu\nu}]_{\theta'}.
\end{eqnarray}
To find the transformation of the action we need to find how the product $F_{\mu\nu}\star_{\theta}
F^{\mu\nu}$ transforms
under (\ref{star-twisted}). For that we use again the deformed Leibniz rule (\ref{modAA}) applied to the
$\theta$-star product of two field strength fields
\begin{eqnarray}
\delta_{\varepsilon}^{\theta'}\Big(F_{\mu\nu}\star_{\theta} F^{\mu\nu}\Big)
&=&\sum_{n=0}^{\infty}{(-i/2)^{n}\over n!}(\theta^{\alpha_{1}\beta_{1}}-\theta'{}^{\alpha_{1}\beta_{1}})
(\theta^{\alpha_{2}\beta_{2}}-\theta'{}^{\alpha_{2}\beta_{2}})
\ldots(\theta^{\alpha_{n}\beta_{n}}-\theta'{}^{\alpha_{n}\beta_{n}})
\nonumber \\
&\times &
\Big\{\Big([\partial_{\alpha_{1}},[\partial_{\alpha_{2}},\ldots[\partial_{\alpha_{n}},\delta_{\varepsilon}^{\theta'}]\ldots]]F_{\mu\nu}\Big)
\star_{\theta}\Big(
\partial_{\beta_{1}}\partial_{\beta_{2}}\ldots
\partial_{\beta_{n}}F^{\mu\nu}\Big)  \label{modFF}\\
& & \,\,\,+\,\,\,\Big(\partial_{\alpha_{1}}\partial_{\alpha_{2}}\ldots
\partial_{\alpha_{n}}F_{\mu\nu}\Big)\star_{\theta}
\Big([\partial_{\beta_{1}},[\partial_{\beta_{2}},\ldots[\partial_{\beta_{n}},\delta_{\varepsilon}^{\theta'}]\ldots]]F^{\mu\nu}\Big)\Big\}.
\nonumber
\end{eqnarray}
Using manipulations similar to the ones applied above, we conclude that the product $F_{\mu\nu}\star_{\theta}F^{\mu\nu}$
transforms as well as an adjoint field
\begin{eqnarray}
\delta_{\varepsilon}^{\theta'}\Big(F_{\mu\nu}\star_{\theta} F^{\mu\nu}\Big)
=[i\varepsilon,F_{\mu\nu}\star_{\theta} F^{\mu\nu}]_{\theta^{'}}.
\end{eqnarray}
In order to finally show that this transformation implies the invariance of the action with 
respect to the star-twisted transformations
(\ref{star-twisted}) we only have to use the cyclic property of the integral with respect to the $\theta'$-star product
\begin{eqnarray}
\delta_{\varepsilon}^{\theta'}S=-{1\over 2g^{2}}
\int d^{d}x\,{\rm tr\,}\Big[i\varepsilon\star_{\theta'}\left(F_{\mu\nu}\star_{\theta}F^{\mu\nu}
\right)-i\left(F_{\mu\nu}\star_{\theta}F^{\mu\nu}\right)\star_{\theta'}\varepsilon\Big]=0.
\end{eqnarray}

With this analysis we have shown how pure noncommutative U($N$) Yang-Mills theories are invariant under star-gauge 
transformations with any value of the noncommutativity parameter provided the Leibniz rule is modified. The extra terms
in the modified Leibniz rule scale with the difference between the two noncommutativity parameters, 
$\theta^{\mu\nu}-\theta'{}^{\mu\nu}$
and therefore vanish for standard star-gauge transformations. In the same way the invariance under the so-called twisted gauge transformations 
of Ref. \cite{vassilevich,twisted_gauge} is recovered when $\theta'{}^{\mu\nu}=0$.

\paragraph{Matter fields.} 
Once studied the case of pure gauge theories, we turn next to gauge field coupled to matter. Because of the peculiar
algebraic properties of noncommutative field theories, the coupling of matter fields to gauge fields can only be made in
 the fundamental, antifundamental and adjoint representation, defined by \cite{hayakawa,no-go}
\begin{eqnarray}
\delta^{\theta'}_{\varepsilon}\psi &=& i\varepsilon\star_{\theta'}\psi \hspace*{1.4cm} \mbox{fundamental,}\nonumber\\
\delta^{\theta'}_{\varepsilon}\psi&=&-i\psi\star_{\theta'}\varepsilon   \hspace*{1cm} \mbox{antifundamental,}
\label{star-twisted-matter}\\
\delta^{\theta'}_{\varepsilon}\psi&=&[i\varepsilon,\psi]_{\theta'} \hspace*{1.4cm} \mbox{adjoint.} \nonumber 
\end{eqnarray}
For standard noncommutative gauge theories the matter action is build in terms of the matter fields and the corresponding
covariant derivatives
\begin{eqnarray}
\nabla_{\mu}\psi &=& \partial_{\mu}\psi-iA_{\mu}\star_{\theta}\psi \hspace*{1.4cm} \mbox{fundamental,} \nonumber \\
\nabla_{\mu}\psi &=& \partial_{\mu}\psi+i\psi\star_{\theta}A_{\mu} \hspace*{1.4cm} \mbox{antifundamental,} 
\label{covariant_derivatives}\\
\nabla_{\mu}\psi &=& \partial_{\mu}\psi-i[A_{\mu},\psi]_{\theta} \hspace*{1.4cm} \mbox{adjoint.} \nonumber 
\end{eqnarray}
These derivatives, by construction, transform covariantly under standard star-gauge transformations, thus 
guaranteeing the invariance of the total action under this symmetry. The interesting point is that 
the previous derivatives can also transform covariantly under (\ref{star-twisted-matter}) provided we use 
the following modified Leibniz rule [cf. Eq. (\ref{modAA})]
\begin{eqnarray}
\delta_{\varepsilon}^{\theta'}\Big(\Phi_{1}\star_{\theta}\Phi_{2}\Big)
&=&\sum_{n=0}^{\infty}{(-i/2)^{n}\over n!}(\theta^{\alpha_{1}\beta_{1}}-\theta'{}^{\alpha_{1}\beta_{1}})
(\theta^{\alpha_{2}\beta_{2}}-\theta'{}^{\alpha_{2}\beta_{2}})\ldots(\theta^{\alpha_{n}\beta_{n}}-\theta'{}^{\alpha_{n}\beta_{n}})
\label{modified_general}
\nonumber \\
&\times &
\Big\{\Big([\partial_{\alpha_{1}},[\partial_{\alpha_{2}},\ldots[\partial_{\alpha_{n}},\delta_{\varepsilon}^{\theta'}]\ldots]]\Phi_{1}\Big)
\star_{\theta}\Big(
\partial_{\beta_{1}}\partial_{\beta_{2}}\ldots
\partial_{\beta_{n}}\Phi_{2}\Big)  \\
& & \,\,\,+\,\,\,\Big(\partial_{\alpha_{1}}\partial_{\alpha_{2}}\ldots
\partial_{\alpha_{n}}\Phi_{1}\Big)\star_{\theta}
\Big([\partial_{\beta_{1}},[\partial_{\beta_{2}},\ldots[\partial_{\beta_{n}},\delta_{\varepsilon}^{\theta'}]\ldots]]\Phi_{2}\Big)\Big\}.
\nonumber
\end{eqnarray}
Indeed, using this expression and with the help of Eqs. (\ref{commutators}) and (\ref{crucial_identity}) we arrive at the
following transformation of the covariant derivatives under the transformations (\ref{star-twisted}) and (\ref{star-twisted-matter})
\begin{eqnarray}
\delta_{\varepsilon}^{\theta'}\nabla_{\mu}\psi &=& i\varepsilon\star_{\theta'}(\nabla_{\mu}\psi) \hspace*{1.3cm}\mbox{fundamental,} 
\nonumber\\
\delta_{\varepsilon}^{\theta'}\nabla_{\mu}\psi &=& -i(\nabla_{\mu}\psi)\star_{\theta'}\varepsilon \hspace*{1cm} \mbox{antifundamental,} \\
\delta_{\varepsilon}^{\theta'}\nabla_{\mu}\psi &=&i[\varepsilon,\nabla_{\mu}\psi]_{\theta'} \hspace*{1.6cm}\mbox{adjoint,}
\nonumber 
\end{eqnarray}
i.e., the covariant derivatives (\ref{covariant_derivatives}) remain covariant under the new transformations if we modify 
the Leibniz rule according to (\ref{modified_general}). 
Notice that the key point here is that all covariant derivatives are defined using the $\theta$-star product $\star_{\theta}$, whereas 
the gauge trasformations involve the $\theta'$-star product $\star_{\theta'}$. 

These transformations of the covariant derivatives under star-twisted gauge transformations 
show that any action constructed in terms of them, invariant under U($N$) $\theta$-star gauge transformations, 
is at the same 
time invariant under $\theta'$-star-twisted gauge transformations for any value of $\theta'{}^{\mu\nu}$. 

\section{Mathematical construction}

After the heuristic construction of star-twisted gauge theories presented above we turn to a more mathematical derivation of these class of
invariances. In the following we use the notation of Ref. \cite{twisted_gauge} with minimal changes. A comprehensive study of the 
Lie algebra of star-gauge transformations can be found in \cite{star-gauge}.

As usual we work with an algebra of functions $\mathcal{A}$ where the standard pointwise product is defined through the operation 
$\mu:\mathcal{A}\otimes \mathcal{A}\rightarrow \mathcal{A}$ as $f\cdot g=\mu(f\otimes g)$. In terms of this, 
the $\theta$-star product is defined
as $f\star_{\theta}g=\mu[\mathcal{F}_{\theta}^{-1}(f\otimes g)]$ where the twist operator is given by
\begin{eqnarray}
\mathcal{F}_{\theta}=e^{-{i\over 2}\theta^{\mu\nu}\partial_{\mu}\otimes\partial_{\nu}}.
\label{twist}
\end{eqnarray}
Given a function $f$ in the algebra $\mathcal{A}$
we define the differential operator
\begin{eqnarray}
X_{f}^{\theta}\equiv \sum_{n=0}^{\infty}{(-i/2)^{n}\over n!}\theta^{\mu_{1}\nu_{1}}\ldots \theta^{\mu_{n}\nu_{n}}
\partial_{\mu_{1}}\ldots\partial_{\mu_{n}}f \partial_{\nu_{1}}\ldots \partial_{\nu_{n}}.
\end{eqnarray}
These operators act on the elements of the same algebra of functions. In particular we can define the left $\theta'$-action 
of this operator on a function $g\in\mathcal{A}$ as
\begin{eqnarray}
X_{f}^{\theta}\rhd_{\theta'}g\equiv \sum_{n=0}^{\infty}{(-i/2)^{n}\over n!}\theta^{\mu_{1}\nu_{1}}\ldots \theta^{\mu_{n}\nu_{n}}
(\partial_{\mu_{1}}\ldots\partial_{\mu_{n}}f)\star_{\theta'} (\partial_{\nu_{1}}\ldots \partial_{\nu_{n}}g),
\end{eqnarray}
which using Eq. (\ref{crucial_identity}) leads to the identity
\begin{eqnarray}
X_{f}^{\theta}\rhd_{\theta'}g=f\star_{\theta'-\theta}g.
\end{eqnarray}
Using this expression, together with the associativity of the star-product, it is easy to show that 
these differential operators satisfy the composition rule
\begin{eqnarray}
(X_{f}^{\theta}\circ_{\theta'}X_{g}^{\theta})\rhd_{\theta'}h\equiv X_{f}^{\theta}\rhd_{\theta'}(X_{g}^{\theta}\rhd_{\theta'}h)
=X_{f\star_{\theta'-\theta}g}^{\theta}\rhd_{\theta'}h
\label{composition}
\end{eqnarray}
for every $f,g,h\in\mathcal{A}$.

The right action of the differential operator $X_{f}^{\theta}$ can be also defined in an analogous way by
\begin{eqnarray}
g\lhd_{\theta'}X^{\theta}_{f}&=&\sum_{n=0}^{\infty}{(-i/2)^{n}\over n!}\theta^{\mu_{1}\nu_{1}}\ldots \theta^{\mu_{n}\nu_{n}}
(\partial_{\mu_{1}}\ldots\partial_{\mu_{n}}g)\star_{\theta'} (\partial_{\nu_{1}}\ldots \partial_{\nu_{n}}f) \nonumber \\
&=& g\star_{\theta'-\theta}f.
\end{eqnarray}
Finally, we define the adjoint action by 
\begin{eqnarray}
{\rm Adj}(X_{f}^{\theta})\rhd_{\theta'}g&=&\sum_{n=0}^{\infty}{(-i/2)^{n}\over n!}\theta^{\mu_{1}\nu_{1}}\ldots\theta^{\mu_{n}\nu_{n}}
\Big[\partial_{\mu_{1}}\ldots\partial_{\mu_{n}}f,\partial_{\nu_{1}}\ldots\partial_{\nu_{n}}g\Big]_{\theta'} \nonumber \\
&=& [f,g]_{\theta'-\theta}.
\end{eqnarray}
Using again the definitions and the associativity of the product it can shown that both the antifundamental and adjoint actions of the differential operator $X_{f}^{\theta}$ satisfy the analog of the composition rule (\ref{composition}). 

The transformations properties of matter fields under $\theta'$-star gauge transformations in the (anti)fundamental and 
adjoint representations introduced above can be written in terms of the action of the operator $X_{f}^{\theta}$ as
\begin{eqnarray}
\delta^{\theta'}_{\varepsilon}\Phi &=& iX^{\theta}_{\varepsilon^{a}T^{a}}\rhd_{\theta+\theta'}\Phi \hspace*{1.85cm} \mbox{fundamental},
\nonumber \\
\delta^{\theta'}_{\varepsilon}\Phi &=& -i\Phi\lhd_{\theta+\theta'}X^{\theta}_{\varepsilon^{a}T^{a}} \hspace*{1.5cm} 
\mbox{antifundamental},
\label{transformations_fields_m}\\
\delta^{\theta'}_{\varepsilon}\Phi &=& i{\rm Adj}(X_{\varepsilon^{a}T^{a}}^{\theta})\rhd_{\theta+\theta'}\Phi 
\hspace*{0.85cm} \mbox{adjoint} 
\nonumber
\end{eqnarray}

Once the action of the transformations are defined on fields we need to extend it to their $\theta$-star products. To this end, 
let us consider two fields $\Phi_{1}$, $\Phi_{2}$ transforming in some of the above representation 
under $\theta$-star gauge transformations ($\theta'{}^{\mu\nu}=\theta^{\mu\nu}$), 
but such that their $\theta$-star product itself transform also in one of these representations 
under the same transformation. This is the case of the product of two 
fields in the adjoint representation, an antifundamental with a fundamental, an adjoint field with a fundamental field or an antifundamental
field with a field in the adjoint representation. We are going to focus on these cases since they are the building blocks
in terms of which the action is constructed.

For concreteness we consider the product of two adjoint fields, although all other cases can be treated in a similar way with the
same result. For star-gauge transformations, $\theta'{}^{\mu\nu}=\theta^{\mu\nu}$, we have that
\begin{eqnarray}
\delta^{\theta}_{\varepsilon}(\Phi_{1}\star_{\theta}\Phi_{2})&=&i[\varepsilon^{a}T^{a},\Phi_{1}]_{\theta}\star_{\theta}\Phi_{2}+
i\Phi_{1}\star_{\theta}[\varepsilon^{a}T^{a},\Phi_{2}]_{\theta} \nonumber \\
&=&i\Big[\varepsilon^{a}T^{a},\Phi_{1}\star_{\theta}\Phi_{2}\Big]_{\theta}\,\,\,,
\label{adjoint_fields}
\end{eqnarray}
so the product of two adjoint fields transforms is an adjoint field itself with the standard coproduct
\begin{eqnarray}
\delta^{\theta}_{\varepsilon}(\Phi_{1}\star_{\theta}\Phi_{2})=\mu\Big[\mathcal{F}_{\theta}^{-1}\Delta(\delta_{\varepsilon}^{\theta})
\Phi_{1}\otimes\Phi_{2}\Big],\hspace*{1cm}
\Delta(\delta^{\theta}_{\varepsilon})=\delta^{\theta}_{\varepsilon}\otimes\mathbf{1}+\mathbf{1}\otimes \delta^{\theta}_{\varepsilon}.
\end{eqnarray}

Equation (\ref{adjoint_fields}) ceases to be valid if the gauge transformations act with a noncommutativity parameter different from the
one used to multiply fields, i.e. $\theta'{}^{\mu\nu}\neq \theta^{\mu\nu}$. In order to keep the covariance of the product of the
two fields we require the transformation
\begin{eqnarray}
\delta^{\theta'}_{\varepsilon}(\Phi_{1}\star_{\theta}\Phi_{2})&=&i{\rm Adj}(X^{\theta}_{\varepsilon^{a}T^{a}})\rhd_{\theta+\theta'}(\Phi_{1}
\star_{\theta}\Phi_{2})\nonumber \\
&=& i\Big[\varepsilon^{a}T^{a},\Phi_{1}\star_{\theta}\Phi_{2}\Big]_{\theta'}.
\label{covariant_product}
\end{eqnarray}
Notice that now we cannot apply the same manipulations used in Eq. (\ref{adjoint_fields}) since in this case we have two different
star products. To simplify the expression we use Eq. (\ref{crucial_identity}) to rewrite it in terms of $\theta'$-star products alone.
\begin{eqnarray}
\delta_{\varepsilon}^{\theta'}(\Phi_{1}\star_{\theta}\Phi_{2})  \hspace*{14cm} \\ 
=\sum_{n=0}^{\infty}{(-i/2)^{n}\over n!}(\theta'{}^{\mu_{1}\nu_{1}}-
\theta^{\mu_{1}\nu_{1}})\ldots (\theta'{}^{\mu_{n}\nu_{n}}-\theta^{\mu_{n}\nu_{n}})\left[i\varepsilon^{a}T^{a},(\partial_{\mu_{1}}\ldots
\partial_{\mu_{n}}\Phi_{1})\star_{\theta'}(\partial_{\nu_{1}}\ldots\partial_{\nu_{n}}\Phi_{2})\right]_{\theta'}.
\nonumber 
\end{eqnarray}
In this way we have that all products on the right-hand side of the previous equation are identical and we 
can use the standard identities for commutators. Applying Eq. (\ref{commutators}) and once again the relation 
(\ref{crucial_identity}) we find that the transformation of the product of the two fields given in Eq. (\ref{covariant_product}) 
can be expressed after some manipulations as
\begin{eqnarray}
\delta_{\varepsilon}^{\theta'}(\Phi_{1}\star_{\theta}\Phi_{2})  
&=& \sum_{n=0}^{\infty}{(-i/2)^{n}\over n!}(\theta^{\mu_{1}\nu_{1}}-
\theta'{}^{\mu_{1}\nu_{1}}) (\theta^{\mu_{2}\nu_{2}}-\theta'{}^{\mu_{2}\nu_{2}})\ldots 
(\theta^{\mu_{n}\nu_{n}}-\theta'{}^{\mu_{n}\nu_{n}}) \nonumber \\ 
&\times &  
\mu\Big\{\mathcal{F}_{\theta}^{-1}
\Big[[\partial_{\mu_{1}},[\partial_{\mu_{2}},\ldots[\partial_{\mu_{n}},i{\rm Adj}(X^{\theta'}_{\varepsilon^{a}T^{a}})]\ldots]] \otimes 
\partial_{\nu_{1}}\ldots\partial_{\nu_{n}}
\label{suggestive2} \\
& &  +\,\,\,\partial_{\mu_{1}}\ldots
\partial_{\mu_{n}}\otimes 
[\partial_{\nu_{1}},[\partial_{\nu_{2}},\ldots[\partial_{\nu_{n}},i{\rm Adj}(X^{\theta'}_{\varepsilon^{a}T^{a}})]\ldots]]\Big]\rhd_{\theta+
\theta'}
(\Phi_{1}\otimes\Phi_{2})\Big\} .
\nonumber 
\end{eqnarray}
Interestingly, this expression can be written in a more compact form 
\begin{eqnarray}
\delta_{\varepsilon}^{\theta'}(\Phi_{1}\star_{\theta}\Phi_{2}) =\mu\Big[\mathcal{F}^{-1}_{\theta}
\Delta(\delta^{\theta'}_{\varepsilon})_{\theta-\theta'}
(\Phi_{1}\otimes\Phi_{2})\Big],
\end{eqnarray}
where the twisted coproduct $\Delta(\delta^{\theta'}_{\varepsilon})_{\theta-\theta'}$ is given by
\begin{eqnarray}
\Delta(\delta^{\theta'}_{\varepsilon})_{\theta-\theta'}=\mathcal{F}_{\theta-\theta'}\Big(\delta^{\theta'}_{\varepsilon}\otimes\mathbf{1}+
\mathbf{1}\otimes \delta^{\theta'}_{\varepsilon}\Big)\mathcal{F}_{\theta-\theta'}^{-1}.
\label{twisted_coproduct}
\end{eqnarray}
For $\theta'{}^{\mu\nu}=0$ we recover the standard twisted coproduct of Ref. \cite{vassilevich,twisted_gauge}, whereas for $\theta'{}^{\mu\nu}=
\theta^{\mu\nu}$ the twist vanishes and we find the standard coproduct associated with the ordinary Leibniz rule.

The same analysis presented here can be repeated for the $\theta$-star products of fields in the representations mentioned above. 
In all cases the requirement that the product
transforms covariantly under $\theta'$-star gauge transformations lead to the same twisted coproduct (\ref{twisted_coproduct}).

\section{Discussion}

In this note we have proved that U($N$) noncommutative gauge theories admit a continuous family of twisted gauge invariances. The two cases 
studied so far in the literature, namely star-gauge symmetry and twisted gauge symmetry, are just two special cases of this more general 
start-twisted gauge invariances. Hence this provides a generalization of the twisted gauge theories proposed
in \cite{vassilevich,twisted_gauge}. 

In order to make things simpler here we have restricted our attention to U($N$) gauge theories. This is the only group for which both star-gauge 
symmetries ($\theta'{}^{\mu\nu}=\theta^{\mu\nu}$) 
and twisted gauge symmetries ($\theta'{}^{\mu\nu}=0$) are consistent with the gauge fields taking values in the Lie algebra
of the gauge group. For any other group the gauge fields have to be valued in the universal enveloping algebra of the corresponding
Lie algebra. This is forced by gauge transformations or the equations of motion respectively. 

Following the analysis of Ref. \cite{LAG-M-VM} one can interpret the extra terms in the twisted Leibniz rule
(\ref{modified_general}) as the result of the transformation
of the $\theta$-star product under $\theta'$-star gauge transformations. In this sense all these star-twisted
invariances of the action are not 
standard symmetries, leading to the usual difficulties in defining Noether currents and Ward identities. There is,
however, an exceptional case in this family of invariances
corresponding to star-gauge transformations ($\theta^{\mu\nu}=\theta'{}^{\mu\nu}$) in which case
one deals with a standard star-gauge symmetry
that acts on the field alone. 
Associated with this symmetry it is possible to derive the corresponding Noether currents and Ward identities
in the usual way. Because of this, 
it can be said that this special invariance, which is a true standard symmetry, plays again a custodial 
r\^ole. To be more concrete, by taking $\theta{}'^{\mu\nu}=\lambda\theta^{\mu\nu}$, with $0\leq \lambda \leq 1$, we find that 
the U($N$) gauge theory action is invariant under a family of 
star-twisted transformations that continuously interpolate between star-gauge symmetry and twisted gauge invariance. Among
them only the point $\lambda=1$ corresponds to a standard gauge symmetry.

We look now at the extreme case $\theta^{\mu\nu}=0$ with $\theta'{}^{\mu\nu}$ nonvanishing, i.e., the
case of a U($N$) = U(1) $\times$ SU($N$) commutative gauge theory. In the pure gauge theory the U(1) part decouples 
from the SU($N$) factors and remains free. However both factors are mixed by 
(\ref{star-twisted}). This means that the terms in the action coupling the U(1) factor to SU($N$), although being zero have 
nevertheless a nonvanishing
variation under these transformations. Actually, taking into account the variation under $\theta'$-twisted
transformations of these couplings is crucial 
to ensure the invariance of the full action.
As in the case of noncommutative gauge theories, the deformed Leibniz rule
in this case can be understood as the result of transforming the standard commutative product in the action. 
Notice, however, that for all we know about nonabelian gauge theories this twisted invariance does not seem to have
any dynamical meaning in the theory. 

We have seen that the class of twisted invariances in noncommutative gauge theories are larger than the so-far studied
twisted gauge symmetry. The existence of this large family of twisted invariances can be interpreted as shedding a
rather pessimistic view on the dynamical significance of these transformations. It seems once more than the 
dynamically relevant symmetry is provided by standard star-gauge transformations, while twisted gauge invariances 
would be accidental.

Before closing, it is important to stress that the analysis presented here cannot be extended to 
space-time symmetries \cite{space-time}. The reason is that, unlike with gauge transformations, we lack a deformed
version of the Poincar\'e algebra. In this sense twisted Poincar\'e transformations seem to play 
a more special r\^ole than twisted gauge transformations.

\section*{Acknowledgments}

We would like to thank Luis \'Alvarez-Gaum\'e and Nicola Caporaso for enlightening  
discussions. The work of A.D.-V. has been supported by a Castilla y Le\'on Regional Government 
Predoctoral Fellowship and by Spanish Science Ministry Grant FIS2006-05319.
M.A.V.-M. acknowledges the support from Spanish Science
Ministry Grants PA2005-04823, FIS2006-05319, Basque Government Grant IT-357-07 and the Spanish Consolider-Ingenio 2010 Program CPAN
(CSD2007-00042), and thanks the CERN
Theory Unit for hospitality during the completion of this work. 

\section*{Appendix}
\renewcommand{\thesection}{A}

In this Appendix we give a brief proof of Eq. (\ref{crucial_identity}). This is based on the following property of the 
twist operator (\ref{twist}) valid for any $\theta'{}^{\mu\nu}$
\begin{eqnarray}
\mathcal{F}_{\theta}^{-1}=\mathcal{F}_{\theta'}^{-1}\mathcal{F}_{\theta-\theta'}^{-1},
\end{eqnarray}
that allows to write the star product of two functions $f(x)$ and $g(x)$ as
\begin{eqnarray}
f\star_{\theta} g=\mu\Big(\mathcal{F}_{\theta}^{-1}f\otimes g\Big)=\mu\Big(\mathcal{F}_{\theta'}^{-1}\mathcal{F}_{\theta-\theta'}^{-1}
f\otimes g\Big).
\end{eqnarray}
By expanding now explicitely the second twist operator inside the bracket and using the linearity of the $\mu$-operation we find
\begin{eqnarray}
f\star_{\theta}g=\sum_{n=0}^{\infty}{(i/2)^{n}\over n!}(\theta^{\alpha_{1}\beta_{1}}-\theta'{}^{\alpha_{1}\beta_{1}})\ldots
(\theta^{\alpha_{n}\beta_{n}}-\theta'{}^{\alpha_{n}\beta_{n}})\mu\Big(\mathcal{F}_{\theta'}^{-1}\partial_{\mu_{1}}\ldots\partial_{\mu_{n}}f
\otimes \partial_{\nu_{1}}\ldots\partial_{\nu_{n}}g\Big),
\end{eqnarray}
which is exactly Eq. (\ref{crucial_identity}).

\end{document}